Milan M. Ćirković
Suzana Cvetićanin

# BACKWARD CAUSATION, ISOLATION AND THE PURSUIT OF JUSTICE

**Abstract.** The recent operationalization of the famous Newcomb's game by Schmidt (1998) offers an interesting and thought-provoking look at the plausibility of backward causation in a Newtonian universe. Hereby we investigate two details of the Schmidt's scenario which may, at least in principle, invalidate his conclusion in two different domains: one dealing with the issue of Newtonian predictability in specific instance of human actions, and the other stemming from a possible strategy aimed at obviating the anthropically oriented view of backward causation as applied to a judicial and ethical problem posed by a version of the scenario. We conclude that the scenario is at least to be more complex than originally presented in order to remain viable. However, it points to a very deep and delicate question of compatibility of backward causation with the conventional ethical standards.

## 1. Introduction: Schmidt's operationalization of Newcomb's "paradox"

In a recent thought-provoking study, Schmidt (1998) has offered a physically plausible way for realization of the famous Newcomb's "paradox" (or Newcomb's game).[1] As is rather well-known, the crucial step in operationalization of this paradox is construction of a physical model for accurate prediction of the outcome of a human choice between the two simple alternatives. Schmidt's solution is based on the possibility of existence of a collection of microscopic beings ("dwarfs") which should, in principle, be able to predict motions of particles in our macroscopic worlds ("gigantions" in colourful jargon of this scenario) with the same degree of confidence human scientists predict motions of objects in celestial mechanics. Their own microworld is based on a different class of particles, called "tinions". Science and technology of dwarfs' civilization are far ahead of ours own in Schmidt's scenario, and they are able to predict motions in all particles in a human being—the player in the Newcomb's game—exactly 24 hours in advance. If specific physical conditions listed in the Appendix of the Schmidt's paper are satisfied, dwarfs are allegedly able to calculate the outcome of the player's decision sufficiently in advance to put in the second box £0 or £1,000,000, in accordance with the game's propositions. This represents a particular form of physical operationalization of backward causation, since it is implied that the player's choice *prima facie* backward causes the decision of dwarfs whether to put the prize into the second box.

There are two basic issues we would like to investigate in the present paper. The first is the problem of isolation of a physical system in the context of suggested Newtonian world as perceived by Schmidt's dwarfs. The second is connected with the second, sinister version of the Newcomb's game presented by Schmidt, in which death of a sentient being depends on the player's choice. In a suggested trial for murder caused through what is perceived as a backward causation primer, the outcome of the trial should not be known in advance of the standard judicial procedure (according to our usual perceptions of justice). However, using the very same model for backward causation, a skillful lawyer can reduce the entire procedure, and jurors' deliberations in particular, to an absurd.

## 2. Are humans sufficiently isolated?

Even the purely classical predictions in the completely understood system will fail if the system is insufficiently isolated. The level of accuracy of predictions achieved in celestial mechanics which are used as analogy of Schmidt is based on an *prima facie* accidental (or explicable teleologically) physical circumstance: the fact that large bodies of the Solar system are few and easily identifiable, as well as located at large distances from outside perturbers. In the specific case, the average distance to other stars at the Solar system's distance from the Galactic center is of the order of magnitude of 1 pc, that is **at least** 4 orders of magnitude larger than the average distance of bodies in the Solar system upon which the predictive power of Newtonian dynamics is usually tested. Since the gravitational force decreases in the Newtonian approximation as the inverse square of distance, this means that Laplace, Gauss, Lagrange and other pioneers of celestial mechanics have been justified in neglecting those perturbation terms arising from interactions with objects external to the Solar system, as classically understood.[2]

It is obvious that the player in the Newcomb's game can not be considered as an isolated system **in the same way** as the Solar system is. Player is a human being, which constantly interacts with its environment by both long- and short-range forces. Objects in his environment are located at distances which are comparable to his characteristic size, i.e. the average distance between two individual gigantions constituting his body. Therefore, there can be no multipole expansion in which all terms of higher order can be neglected, as it is the standard case in celestial mechanics. Self-consistent solution of the equations of motions for gigantions in his body require not only their complete initial conditions, but also initial conditions of all objects in the player's environment. In Schmidt's story, for instance, the player has a prolonged interaction with gigantions in his bed during the night before the game. However, the initial state of his bed has been determined, by other things, by the motion of the player during previous evening, as well as his tidy/untidy habits, presence of other people in his living space, etc. The motions of the player during the previous evening, in its turn, has been influenced among other things by the quality of his dinner, which has been prepared by a cook in a restaurant, etc. etc. provided that all these interactions have taken place within previous 24 hours. These are only short-ranged interactions. A host of further unpleasant questions are posed by the hardly-fathomable but entirely conceivable influence of long-range forces. Is, for example, our behavior influenced by lunar and solar tides, as has been frequently suggested in the course of human history? And even if one could show that these phenomena are irrelevant to our behavior (which has not happened so far), we need still to be good enough Cartesians to admit that they **do influence** the trajectories of motion of gigantions in our body. And the situation is *per analogiam* the same with terrestrial and extraterrestrial electromagnetic fields, as well as any other, currently unknown (but very plausible, like the universal scalar field beloved in the contemporary field theory and quantum cosmology) long-range interactions.

Here the Newtonian simplicity we got accustomed to betrays us. Since in the Newtonian mechanics there is no limiting velocity of propagation for interactions, the fully self-consistent picture would have to be the Laplacean one—that is, taking into account the state of gigantions in the entire universe. We need not dwell upon various criticisms which this Laplacean determinism has faced; some of the recent ones can be found in the systematic exposition by Feinberg, Lavine and Albert (1992). We need only to ask whether the observations of gigantions in the entire universe (and the Newtonian universe is infinite!) is feasible by however advanced science of dwarfs. It seems clear that the answer is negative. In the work of Feinberg et al. it is emphasized that

> in the case of the future motion of planets in the solar system, for example, it would be necessary to include the effects of other stars, and ultimately of other galaxies, if the predictions are to be extended sufficiently far into the future. If some of these distant influences are omitted, then the predictions will become increasingly less accurate as time goes on. Ultimately, since every mass in the universe can influence every other one at least throught gravity, a precise description of the future motion of any body would have to include the effects of all other bodies.

Practically the identical statement could be used in description of the problem of prediction of human motions from the point of view of dwarfs. Only issue here is the quantitative balance between the short time-scale (24 hours) and much vaster complexity of human motions and far smaller degree of isolation of individual human system under considerations.

However, many additional physically realistic assumptions can be made to alleviate this problem. We may abandon strictly Newtonian physics, but remain in the classical (i.e. non-quantum) framework if we assume that the limiting velocity of interaction is $c$, in accordance with the relativity theory. In that case, the cosmological boundary conditions become irrelevant, and all interesting phenomena need to be observed only in the sphere of radius of 24 light-hours (a few times larger than the Solar system[3]). This is more modest requirement, but, unfortunately, it introduces an additional problem, which can be neglected in the strictly Newtonian case, and that is the problem of information **transfer** to the machine doing actual predictive calculation. Streams of data incoming onto the dwarfs' machine will need a substantial fraction of the time interval of 24 hours to reach the machine, and the data from the boundary of this (conventionally speaking!) causally connected region will arrive exactly 24 hours later, leaving no time whatsoever for calculation (which, as has been discussed in the information theory and cybernetics, needs a finite amount of time under all circumstances). If we assume that you still can calculate **something** with a partial amount of data, we run into a sort of informational Zeno's paradox (of Achilles and the tortoise): more data you have to analyze, lesser time you have for the computations, and lesser accuracy will be obtained. Thus, the perfect accuracy required, supposedly, for predicting the behavior of such a complex system as the player is when faced with the two boxes in the morning, is not achieved.

This objection can be stated in the following precise manner. In his Appendix dealing with the physical side of the proposed scenario, Schmidt puts forward the following necessary conditions:

> In order for Newcomb's game to be possible in the way I described it, it needs to be the case that dwarfs can exist which satisfy the following requirements:
>
> 1. They need to be able to perform observations which are sufficiently accurate and comprehensive that, on their basis, the player's actions can be predicted 24 hours in advance.
> 2. They need to be able to calculate their predictions on the basis of this observational data in a sufficiently short time...

The other two requirements concern interactions between the world of dwarfs and our world, and are hardly questionable whatsoever. However, these two initial requirements warrant, as we have seen, extremely close scrutiny. What we question here needs not be anyone of them individually, but rather the **coexistence** of assumptions 1. and 2. in Schmidt's appendix. While each of these assumptions may be valid independently (and this applies more to the assumption 2, rather than the assumption 1. which is uncertain, in our opinion, under any circumstances[4]), their coextensive temporal validity is seriously doubted by external interactions.

Additional difficulty presents the fact that, if the backwards causation is real, even in the limited sense of Schmidt (1998), than occurences at some point A distant from the player but in (conventional) causal contact with him (i.e. less than 24 light-hours distant) may be determined by future events which will occur in 24-light-hours sphere **around A**. This, in turn, would require additional prediction in the A's region, requiring observations, some of which will have to be performed more than 24 light-hours from the point of location of the player. Although admittedly contrived, there is no obvious reason why this chain of occurences cannot be constructed with some ingenuity and the dwarfs' powerful technology. Thus, we are again left with the cosmological boundary conditions.

### 3. Backward causation and human justice

We now consider the trial of the player in modified, "deadly" version of Newcomb's game. In this version, as told by Schmidt, we have evil tiny robots which are programmed to kill an innocent dwarf if and only if the player in the Newcomb's game is predicted to take the box with the £1,000 cheque. The moral and juridical responsibility of the player for the crime which has occured in the past now becomes an issue. (Of course, as noted by Schmidt, the primary guilt lies with the programmer of the robots.) Thus, Schmidt envisages a trial of the player for complicity in murder in (human, and the importance of this qualification will become clear from our further discussion) court. While the defense tries to exculpate the player by invoking the deterministic arguments according to which his actions could not be essentially different, since his decision was determined by motion of gigantions inside his body, the prosecution bases its case on absence of such a necessity as perceived by the player himself, and his greed which led him to take the money. In the deliberations, two different perspectives are noticed and defined by Schmidt in the following manner:

1. *Mechanically oriented description of the scenario:* The particular configuration (and motion) *C* of the particles in the player's body, in the boxes, and in the environment shortly before the moment when the prediction was made *caused* the robot to predict that the player would take the £1,000. (In consequence, it killed a dwarf.)...
2. *The anthropically-oriented description:* After thinking about the tiny innocent dwarf for a while, which would be killed if and only if the player were to take the £1,000, the greedy player nevertheless decided to take the £1,000, because he wanted the money to buy a new car. *As a result*, this is what the tiny robot predicted (by some technical method which we understand in principle, but in which we have no particular interest), and so it killed the innocent dwarf.

However, only the anthropically-oriented description fits well with the normative concepts of human justice, as well as our intuitions about the world (which do **not** incorporate backward causation!). Accordingly, as we have expected, the court[5] pronounces the player guilty for the complicity in dwarf's murder.[6]

As Schmidt correctly notes, this judgement is in agreement with our common moral judgement. However, it can hardly be the end of the story. While the judgement is **necessarily** based upon compatibilism (of free will with deterministic classical physics), it still can lead upon conclusions we are, to say at least, uncomfortable with. In order to demonstrate this, let us modify the story of the trial in the following way, noticing that the mechanically-oriented description fits better with the defense strategy, than with the prosecution. For that reason, the defense may introduce, in the last moment, a surprising witness: the famous dwarf engineer and programmer of the predicting machine. We may also, in order to improve upon the setting of the story, assume that the engineer is a little bit scientistically-minded and represents a fraction of the dwarven society which is worried that the entire scandal with the murder could adversely influence the project of further development of the predicting machines. The investigation goes as follows (presumably using a particular high-tech device enabling communication between the court and the witness in real time):

<u>Defense</u>: As far as we understand, your Machine is not permanently working?
<u>Engineer</u>: No, of course not. The reasons of energy consumption prevent us from having it permanently predicting behaviour of a selected target 24 hours in advance. That is the reason the Newcomb's game has been set up only in individual **instances**.
<u>Defense</u>: Can you describe for this court the instance of its latest use?
<u>Engineer</u>: With pleasure. While personally believing—together with most of the scientific community of ours (in sad contradistinction to our nation as a whole)—that the defendant is not guilty, as well as motivated by scientific reasons of further exploration of the predictive capabilities of the Machine, we have decided to put it on further test. Therefore, before coming to this courtroom, I have had the latest round of predicting the behavior of the jury members during next 24 hours. I hope you understand that it is quite a feat, much more difficult than in cases of prediction for the Game. However, **it is based on the same physical principles**. Therefore, just as in the case of our predictions of the choice of player in the experiment you humans call the Newcomb's game, we have accurately predicted the outcome of jury's deliberations, i.e. whether the defendant will be found guilty or not. After all, it is a simple binary choice, just as in the original version of the Newcomb's game.
<u>Defense</u>: So, what is your prediction?
<u>Judge</u>: Although there is no relevant formal precedents, I warn you that this may constitute an illegal influence upon the jury.
<u>Engineer</u>: We have envisaged this possibility on the basis of our—still only preliminary—understanding of human concepts of just trial. Therefore, we have put the predicted verdict into a sealed envelope which we shall leave in the possession of the judge.
<u>Judge</u>: And I shall open it only after the jury deliberations are over, and the verdict is pronounced.

What can be outcome of such a trial? From both our colloquial and "metaphysical" points of view, it does look that the only consistent judgement could be pronouncing the player not guilty. To perceive this, assume for the moment that jurors decide that the player is guilty for murder. Dwarf engineers have, evidently, predicted this and therefore after breaking the seal, the judge will read the "guilty" verdict. However, does this not show (*a posteriori*) that the mechanically-oriented description is valid? And that, consequently, the player should have been acquitted. One should recall that the text in the sealed envelope is based **exclusively** on the technically-oriented description (the only one relevant for dwarfs, as we shall discuss further below). By pronouncing the defendant guilty as charged, the jury will offer an unmistakeable proof of the relevance of the technologically-oriented description of circumstances leading to the crime, that is, supporting the assertion of the defense that there was no real alternative for the player than to take the £1,000 box with all consequences. In other words (more of colloquial nature), how are jurors to be supposed to judge upon the fate of a human being, if they are subject to the same necessity as the defendant?

At this point, the prosecution may choose from two strategies we envisage at present. It may emphasize the uncertainties in the prediction (and especially in cases which, like jury's deliberations, **necessarily include strong inter-human interactions**), along the lines we have discussed in §2 and others—that is, essentially try to refute the accuracy of the Machine's prediction in the changed circumstances. The second possible recourse would be to deny the value of this new application of the predicting Machine, since it is still grounded in the mechanically-oriented description, and therefore irrelevant to human justice.

We believe that the dwarven technology will be able to withstand any attack along the prosecution's first strategical option. That is, **under the assumptions implicit in the very formulation of the trial**, that is, the unmistakeable correlation of murder with the subsequent "greedy" choice of the player. If the uncertainties of §2 (or any other **physical** problems) prove fatal for the very concept of accurate prediction machine, so better for the player, since she will never have to face trial.[7] However, it is hard to see why the purely quantitative increase in its task when applied to the entire jury should **necessarily** invalidate the work of the sufficiently advanced prediction machine.

The second strategy left for the prosecution has much better chances of success—at least as far as human jurors, lawmakers and the general public are concerned—but it is very important to fully understand its consequences and the price that comes with it. The prosecution may opt to call upon thinkers of anti-reductionist attitude as witnesses—for instance Prof. Roger Penrose[8] of that world may be called to testify—in order to prove that there is no common ("middle") ground between the mechanically-oriented and the anthropically-oriented descriptions of even the very same sequence of events (even more reproachable from this point of view would be to attempt to derive the anthropically-oriented description from the mechanically-oriented one). The justice has nothing to do with mechanically-oriented description whatsoever; it entails only the anthropically-oriented description by definition. If the anthropically-oriented description can be construed as including a crime, than it is the simple duty of the court to establish it (by actual reconstruction of that description), and acts subsequently to punish the guilt.

In effect, this line of attack is parallel to the (in)famous central compatibilist thesis that although one possesses a capacity to act otherwise, one still will not exercise this capacity. Parallel but not equivalent, because the law is intended to be concerned with acts exclusively, and not just any acts in metaphysical sense, but acts associated with adequate records in the sense of Feinberg et al. (1992). Thus, the second strategy for prosecution when faced with the specific prediction of the outcome of the trial reduces, in our opinion, to a *quid pro quo*. This form of reasoning means that effectively, the trial is not trial of historical actions, but of capacities—and it does seem too high a price.[9]

One needs to note, finally, another off-shoot of the fact that even in the original Schmidt's presentation, the mechanically-oriented description given by the dwarf engineer is basically a defense testimony, which the player's lawyer could invoke to support his opinion on the innocence of the player. On the other hand, the judgement pronounced in Schmidt's story does seem both juridically and ethically sound to us. Does this mean that dwarfs necessarily possess a different juridical and ethical systems? Should we conclude that these issues are **not scale-invariant**? Not at all, since *in stricto sensu* the engineer's testimony (in the original story) only points to a necessary degree of **observer-dependence** of the entire issue. Dwarfs may be complete libertarians in their own (tinion) society, and possess the ethical norms which severely punish both premeditated and accidental murder of a member of any sentient species, irrespectively of size and other secondary properties (as one ought to naturally expect from a truly advanced and enlightened society). However, their advanced science teaches them that **our** world (complete with behavior of us and other macroscopic creatures) is deterministic, at least **to the degree required by their experiments**. There is no incompatibility here, since the deterministic notions apply on the level **outside** of the dwarven society. To the exactly the same degree, **capacities** of dwarfs are different from those of humans. There is, again, no grounds for our possible moral indignation over the attitude of the dwarf engineer or the mechanically-oriented description. As a smart lawyer (or philosopher) among both dwarfs and humans could perceive, what is really in default here is the Schmidt's artificial division of the universe[10] into strictly distinct gigantion and tinion domains.[11] That is, the necessity of some sort of what is usually called truth value relativism arises when the boundaries of the two worlds (gigantion and tinion) are crossed. (And in this world the issue of logical incoherence of truth value relativism in the particular case of dwarf vs. human moral is no more.) This seems like another issue of relation between the properties of agency and the backward causation, which has been recently emphasized by Brown (1992). The basic difference between us and the dwarfs in the Schmidt's story is the one in **agency**. However, this touches the broader problem whether the backward causation is compatible with some or all human ethical standards. We can not enter into discussion of this topic here.

## 4. Conclusions

The appeal to human language explicated by Schmidt in the §2 of his paper is important, but should not be overrated. As we have discussed, the appeal to human language can actually made the situation worse, for while sticking to the formalities, one may condemn the player in the deadly version of the game, one intuitively (and on the colloquial level) can not understand why should this be so when fair trial implies that the outcome is not known before the end of the judicial procedure, and when jurors are supposed to be under no pressure to agree on one or another verdict. Therefore, the norms of human language and the notion following from it (as the very notion of the "fair trial", for instance, which can be hardly labelled as "metaphysical"), may actually increase our difficulties, instead of liberating us from them. This applies to the entire research on causation. As elaborated by Collier (1999), causation represents a transfer of information, either in metaphysical or colloquial domain. Otherwise, the very attempt made by Schmidt to justify the assumptions in his scenario by appeal to postulates of the classical physics is *non sequitur*. The problem of isolation of systems under consideration is clear in both the metaphysical and everyday treatment of causation. If a traffic jam prevented me from attending an important business meeting, and therefore caused a chain of subsequent occurences in my life, it is only natural that I hold all drivers present at streets that day, as well as the city transportation authorities (partially) responsible for the situation. It is only natural that it is concluded without any appeal to "metaphysical" assumptions. We conclude that the problems exposed here may not be avoided by passing-the-buck to the "metaphysical causation".

This is not to say that Schmidt's model does not provoke very interesting issues of rather "metaphysical" nature. The relationship between various physical models of prediction to the notion of backward causation is certainly one of them, deserving further attention. In this respect, an interesting question to be answered by future investigation is whether the classical uncertainties in the Schmidt's model represent a new argument against the backward causation in **any** classical world? However, we can not delve at this rather subtle issue at present.

Another specific property of the model discussed is that this particular form of backward causation is rather exclusive, in the sense that existence of another form of such causation may interfere with the predictive processes in Schmidt's scenario, thus invalidating the accuracy and the very point of the scenario. Suppose, for instance, that some sufficiently powerful and knowledgeable humanitarian organization perceives the disastrous results of a described experiment (the murder of a dwarf and subsequent trial), and decides to prevent this by supplying to the robots a **false report** on machine's prediction *viz.* the player's decision. If they do it **later** and using some physically different model of backward causation, they will invalidate the accuracy of the original model. The report of the machine represents a **record** in the sense of Feinberg et al. (1992), and it clearly limits **all** subsequent retrodictions—for instance those by the police investigators. (This turn of events may, in contradistinction to the Schmidt's model, initiate some well-known causal paradoxes, but this is not important for us here.) Therefore, the Schmidt's model is exclusive enough.

Schmidt's model is certainly an ingenious attempt to use properties of the classical physics to achieve realization of the Newcomb's game, which certainly deserves to be investigated in detail. Although we do not find direct inconsistencies in it, we do warn that its simplicity is misleading; both its physical and its ethical aspects need to be more complicated, to say at least, in order to remain consistent and viable.

**Acknowledgements.** The authors are happy to express his gratitude to Prof. Petar Grujić for bringing our attention to this issue, as well as Maja Bulatović and Vesna Milošević-Zdjelar for invaluable help in obtaining some of the references.

Milan M. Ćirković
*Astronomska Opservatorija*
*Volgina 7, 11000 Belgrade, SERBIA*
*e-mail:* `arioch@eunet.yu`

Suzana Cvetićanin
*Institute "Mihajlo Pupin – Poslovne usluge"*
*Volgina 15, 11000 Belgrade, SERBIA*
*e-mail:* `scvetic@labtel.imp.bg.ac.yu`

---

[1] For the original formulation and ramifications dealing with the decision theory, see Nozick (1969). In it a human being (henceforth "the player") is confronted with the following game set up by a hypothetical being or machine (or collection of beings or machines) capable of predicting accurately the player's actions. The player is shown two boxes, $B_1$ and $B_2$; the box $B_1$ contains £1,000 and $B_2$ contains either £1,000,000 or nothing. Her options are to take either $B_2$ or both boxes; in former case, the predictor has already put in it £1,000,000, and in the latter case there will be nothing in $B_2$, so the player gets only £1,000. Important additional evidence is that the predictor has performed his actions **in advance**, and can not subsequently influence the contents of the boxes. What is the correct choice? A rather comprehensive list of references up to 1987, together with important theological implications is given by Craig (1987).

[2] Moreover, the terms due to interactions with a large number of small bodies in the Solar system are neglected too; otherwise, no progress could be made whatsoever, since we are still busily discovering new members of the Solar system, in recent years notably those of the Kuiper Belt (e.g. Jewitt 1999). However, in order to make the point of argumentation as simple as possible, let us suppose that the dwarfs' inventory of gigantions making up the player of the Newcomb's game is complete, and consider only external influences.

[3] As defined by orbit of the outermost planet, Pluto. It is, however, still deep inside compared to the Oort comet cloud.

[4] Inherent uncertainty here is succintly put by Feinberg, Lavine and Albert (1992):

> It should be also pointed out that the specification of even a single physical quantity to arbitrary precision requires giving a real number, which contains an infinite amount of information, i.e. all of its digits. A more prudent approach to what must be specified in order to make predictions would be to find out how much can be predicted on the basis of a finite amount of information, such as the first N digits of the initial position and velocity of an object.

The issue now is: how big N must be for each gigantion in order for the complex organization of human brain to decide in the case of the Newcomb's game? Intuition suggests that such a complex system is likely to be what Feinberg et al. (1992) call "extremely sensitive to initial conditions", in other words, N must be very large. However, the Schmidt's scenario is not elaborate enough for our conclusion whether tinion-based machinery (which are still limited by basic requirement of finite memories and computing capacities) can make the problem tractable or not.

[5] Or, more precisely, the judge. In the discussion below, we have chosen to have a (petty) jury pronouncing the verdict, because the situation is essentially the same, and the irrelevance of any quantitative increase in the predictive task is better demonstrated.

[6] Although Schmidt does not elaborate on the issue, obviously the programmer of the killer robots bears **at least** the same degree of guilt. This circumstance is not important for our discussion, although it may well be in the context of the wider investigation of judicial and ethical consequences of operationalized backward causation.

[7] That is not to say that the murder still could not occur and be accompanied by some particular action of the player, but we shall have no evidence whatsoever about the causal connection between the two. This is similar to the Mellor's barbecuer argument against the backward causation (see the discussion in Brown 1992).

[8] See, for instance, Penrose (1989).

[9] Not only because this form of anti-reductionism—ironically enough, when the ideological roots are taken into account—clandestinely reintroduces the horrifying notion of "objective guilt", which (implicitly coupled with "subjective innocence") formed the very basis of the Soviet Marxist "judicial" system and is responsible for at least a dozen of million deaths in the greatest mass murder in human history. What people were held responsible for in the Stalinist version of "justice" were actually capacities in the compatibilist sense, certainly not actions.

[10] Understood as the "intelligible whole"; see, for instance, Munitz (1986).

[11] And we may as well notice that the anthropically-oriented description is not called as it is without reason; it is **meant to** apply in the human society exclusively.